\documentclass[%
 reprint,
 superscriptaddress,
 amsmath,amssymb,
 aps,
]{revtex4-2}

\usepackage{graphicx}
\usepackage{dcolumn}
\usepackage{bm}
\usepackage{braket}
\usepackage{color}
\usepackage{hyperref} 

\usepackage{fancyhdr}
\begin{document}
\fancyhead[R]{\ifnum\value{page}<2\relax\else\thepage\fi}

\preprint{APS/123-QED}

\title{Information loss and run time from practical application of quantum data compression}

\author{Saahil Patel}
\email{saahil.patel@us.af.mil}
\affiliation{Air Force Research Laboratory, Rome, NY}

\author{Benjamin Collis}
\affiliation{Air Force Research Laboratory, Rome, NY}
\affiliation{Griffiss Institute, Rome, NY}

\author{William Duong}
\affiliation{Rochester Institute of Technology, Rochester, NY}

\author{Daniel Koch}
\affiliation{Air Force Research Laboratory, Rome, NY}
\affiliation{Griffiss Institute, Rome, NY}

\author{Massimiliano Cutugno}
\affiliation{Air Force Research Laboratory, Rome, NY}

\author{Laura Wessing}
\affiliation{Air Force Research Laboratory, Rome, NY}

\author{Paul Alsing}
\affiliation{Air Force Research Laboratory, Rome, NY}

\date{\today}

\begin{abstract}
We examine information loss, resource costs, and run time from practical application of quantum data compression. Compressing quantum data to fewer qubits enables efficient use of resources, as well as applications for quantum communication and denoising. In this context, we provide a description of the quantum and classical components of the hybrid quantum autoencoder algorithm, implemented using IBM’s Qiskit language. Utilizing our own data sets, we encode bitmap images as quantum superposition states, which correspond to linearly independent vectors with density matrices of discrete values. We successfully compress this data with near-lossless compression using simulation, and then run our algorithm on an IBMQ quantum chip. We describe conditions and run times for compressing our data on quantum devices.

\end{abstract}

\maketitle

\thispagestyle{fancy}

\section{\label{sec:level1}Introduction}

In the Information Age \cite{castells}, data compression has become a powerful tool for dealing with datasets large in number, volume, and variability. Reducing the dimensionality of large datasets allows for efficient use of resources, as well as extraction of useful features. Classical autoencoders are a type of artificial neural network used extensively for feature extraction \cite{meng, vincent2008}, denoising \cite{lecun, vincent2010}, and data compression or dimensionality reduction\cite{bourlard, hinton, gallinari}.

Classical autoencoders, as well as variational autoencoders \cite{kingma}, can be utilized to help mitigate some of the limitations we find in quantum computing \cite{khoshaman}. With small numbers of noisy qubits in current quantum chips, implementation of quantum algorithms can benefit greatly from efficient use of qubits for data encoding and computation. Quantum data compression has been studied extensively as a means to efficiently use qubits \cite{plesch} and for denoising quantum states \cite{bondarenko}. We can also utilize quantum data compression in the field of quantum communication. Compressed quantum data can be sent between two parties on a public channel, with a private key used to compress or decompress the data, similar to sending classical compressed data over a network \cite{wilde}.

The theoretical limit to losslessly compressing quantum information down to a latent space requires the number of maximum linearly independent vectors from the input state to not exceed the size of the latent space \cite{huang}. In practice however, even if we meet this criteria, we find that the performance and efficiency of quantum data compression is limited by the choice of parametric circuit, the classical optimizer, and the gate depth of the circuit. 

In this study, we investigate information loss in application of quantum data compression utilizing the quantum autoencoder as formulated in Romero et al. (2017) \cite{romero}. 
We study the data compression performance of three parametric circuits on a quantum simulator, and compare the times required to train the network. We then look at a simpler quantum data compression scheme to implement on real qubits that can yield a reasonable training time, with results subject to noise from the quantum devices. 
Because of long training times necessary for quantum data compression, results from a real device face considerable information loss from noise when compared to simulations. Thus, reduction of gate depth can help mitigate expensive resource costs. We test the parameterized circuits on an IBMQ quantum device \cite{ibmq} in order to understand the running time of the data compression training network when compared to simulations. 

In total, we study three parameterized circuits, and compare their perform against measures of expressibility and entangling capability as described in Sim et al. (2019) \cite{sim}. Our simulations demonstrate near-lossless compression of quantum data using an image dataset, where each image is encoded as a unique superposition state.

The layout of this paper is as follows: Section II covers the data compression method, which includes the general Quantum Autoencoder framework, the parameterized quantum circuits, and the optimizer we use for this study. Section III covers our dataset and data encoding for the simulation and quantum chip. Section IV covers results from all our simulations and experiments, and discusses the results. In Section V we summarize the paper and provide future prospects for this research.

\section{Method}
\subsection{Classical Autoencoder}
\begin{figure}
\includegraphics[width=0.45\textwidth]{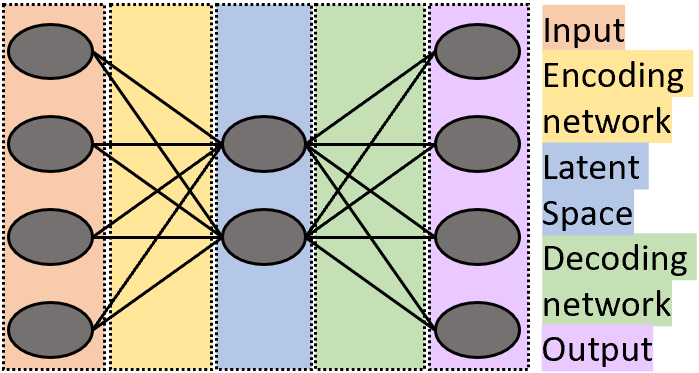}
\caption{The classical autoencoder schematic broken into 5 major components; we start with 4 nodes, compress down to 2 nodes, and decompress back to 4 nodes.}
\label{fig:classical-ae}
\end{figure}

An autoencoder is an artificial neural network that is used to learn the representation (or features) of an unlabeled dataset. The encoder improves and validates the representation by regenerating the input. It is a dimensionality reduction algorithm that takes in data and compresses it to a latent space. We can think of the latent space as containing the ``essence'' of the input data - a compressed representation where similar data points are closer together in space. The decoding network then extracts features from the latent space and reconstructs the image, ideally without any noise or unwanted artifacts. A full autoencoder network consists of an encoding neural network, a latent space, and a decoding network, as shown in Figure~\ref{fig:classical-ae}. 

\subsection{Quantum Autoencoder}
\begin{figure}
\includegraphics[width=0.5\textwidth]{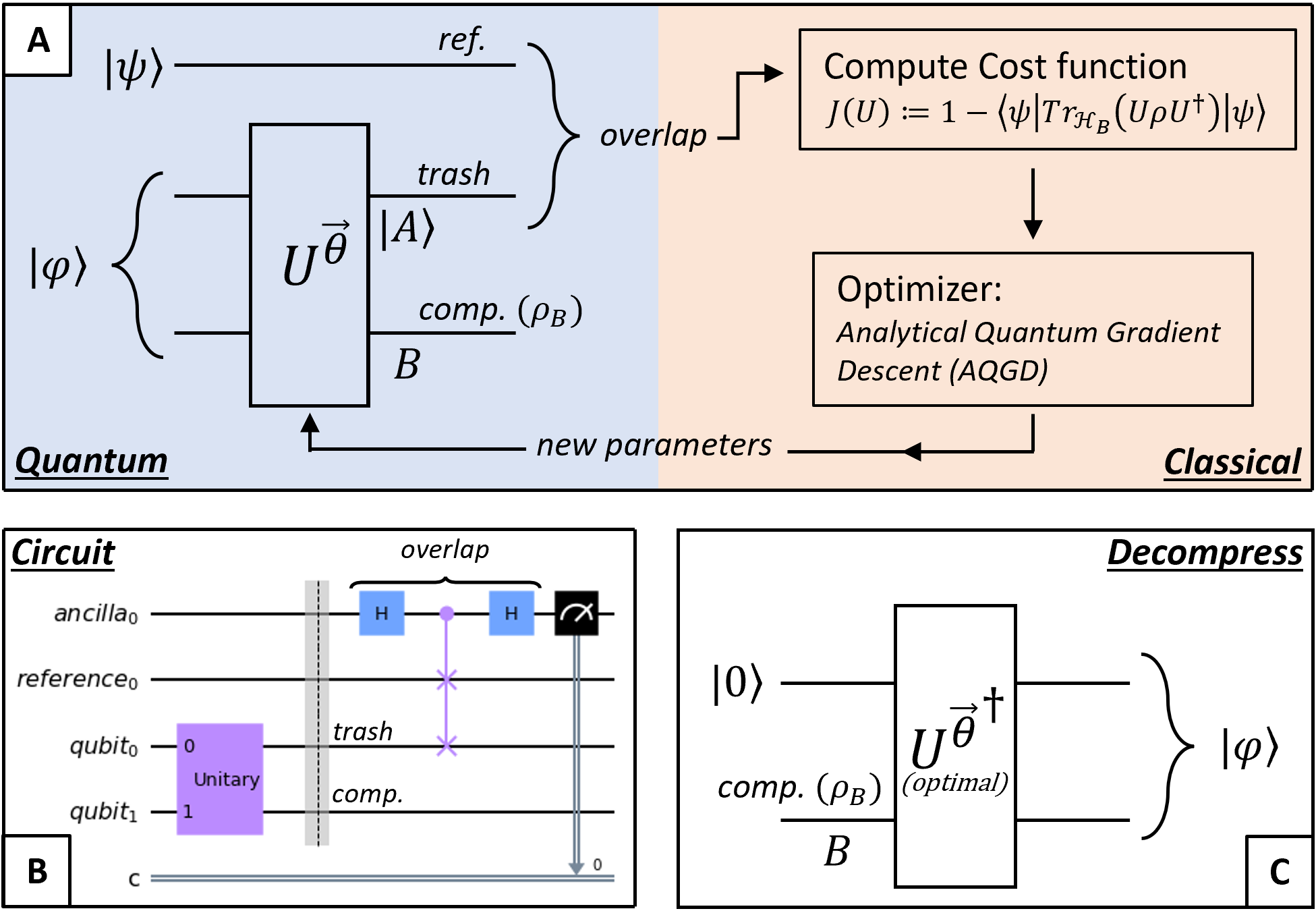}
\caption{\textbf{A}: The QAE schematic with the quantum component on the left containing the input state, PQC, and the overlap. The classical component is on the right containing the computation of the cost function and the optimizer. \textbf{B}: The Qiskit circuit for the quantum subroutine, showing the Swap Test to compute the overlap. \textbf{C}: Decompressing the compressed quantum state by bringing in a new qubit, and using optimal parameters for lossless decompression.}
\label{fig:qae-schematic}
\end{figure}

The quantum autoencoder (QAE) is a hybrid algorithm, where the encoding and decoding neural networks of a classical autoencoder are replaced by quantum subroutines that use quantum data, utilizing parameterized unitary gates and entanglement. The circuit's aim is to compress information from an input state $\ket{\varphi}$ to a mixed compressed state $\rho_B$, labeled in Figure~\ref{fig:qae-schematic}A as $B$. With optimal parameters $\vec{\theta}$, we can get lossless compression, where all the information from $\ket{\varphi}$ is compressed down into state $\rho_B$. The remaining ``trash" state is an empty state ($\ket{0}^{{\otimes}n}$), shown as $\ket{A}$ in Figure~\ref{fig:qae-schematic}A.
To do this, the overlap between the trash state and a reference $\ket{0}^{{\otimes}n}$ state is computed, shown as $\ket{\psi}$ in Figure~\ref{fig:qae-schematic}A. Extracting the trash state $\ket{A}$ requires taking the partial trace on the mixed state $U{\rho}U^\dagger$ with respect to $B$, where $U$ is our parameterized circuit, and $\rho=\ket{\varphi}\bra{\varphi}$. The overlap, $\bra{\psi}{Tr_B}(U{\rho}U^\dagger)\ket{\psi}$, is a measure of fidelity between $\ket{\psi}$ and $\ket{A}$. The maximum of this fidelity measure is 1, which ensures that the trash state is empty, and the compression is lossless. So the task of the classical optimizer is to minimize the cost function $J(U)$, defined as
\begin{eqnarray}                 
	J(U)=1-\bra{\psi}{Tr_B}(U{\rho}U^\dagger)\ket{\psi}.
	\label{cost}
\end{eqnarray} 
Within the quantum circuit, the overlap between the reference ($\ket{\psi}$) and trash ($\ket{A}$) states is computed using the Swap Test \cite{buhrman}, as shown in Figure~\ref{fig:qae-schematic}B. The Swap Test is a quantum operation that measures the difference between two quantum states. From Figure~\ref{fig:qae-schematic}B, we measure out the ancilla qubit, which computes the overlap between the two states, reference and trash, using a control-swap gate. If the two states are orthogonal, then the probability of the ancilla qubit measured as $\ket{0}$ is $1/2$ ($J(U)=1/2$), while if the two states are the same, the probability of measuring the ancilla as $\ket{0}$ is $1$ ($J(U)=0$)
\cite{schuld0}.

Once the training circuit minimizes the cost function, we get the final set of parameters $\vec{\theta}$ that compresses the dataset as efficiently as the parameterized quantum circuit (PQC) allows. The performance of the PQC is measured by two circuit descriptors, entangling capability and expressibility \cite{sim}, discussed in the next section. The success of the algorithm is gauged by using new qubits to decompress the latent space back to the original state. Since we are gauging the efficiency of the compression, the decompression is not a training network. We simply use the optimal set of parameters to implement $U^\dagger$ on the qubits, as shown in Figure~\ref{fig:qae-schematic}C. The optimal parameters take in new qubits and the compressed state $\rho_B$, and decompresses the data back to the original input state $\ket{\varphi}$. For this study, we measure the fidelity between the original input and the final decompressed output to understand information loss and compression efficiency.

\subsection{Parameterized Quantum Circuit}
\begin{figure}
\includegraphics[width=0.5\textwidth]{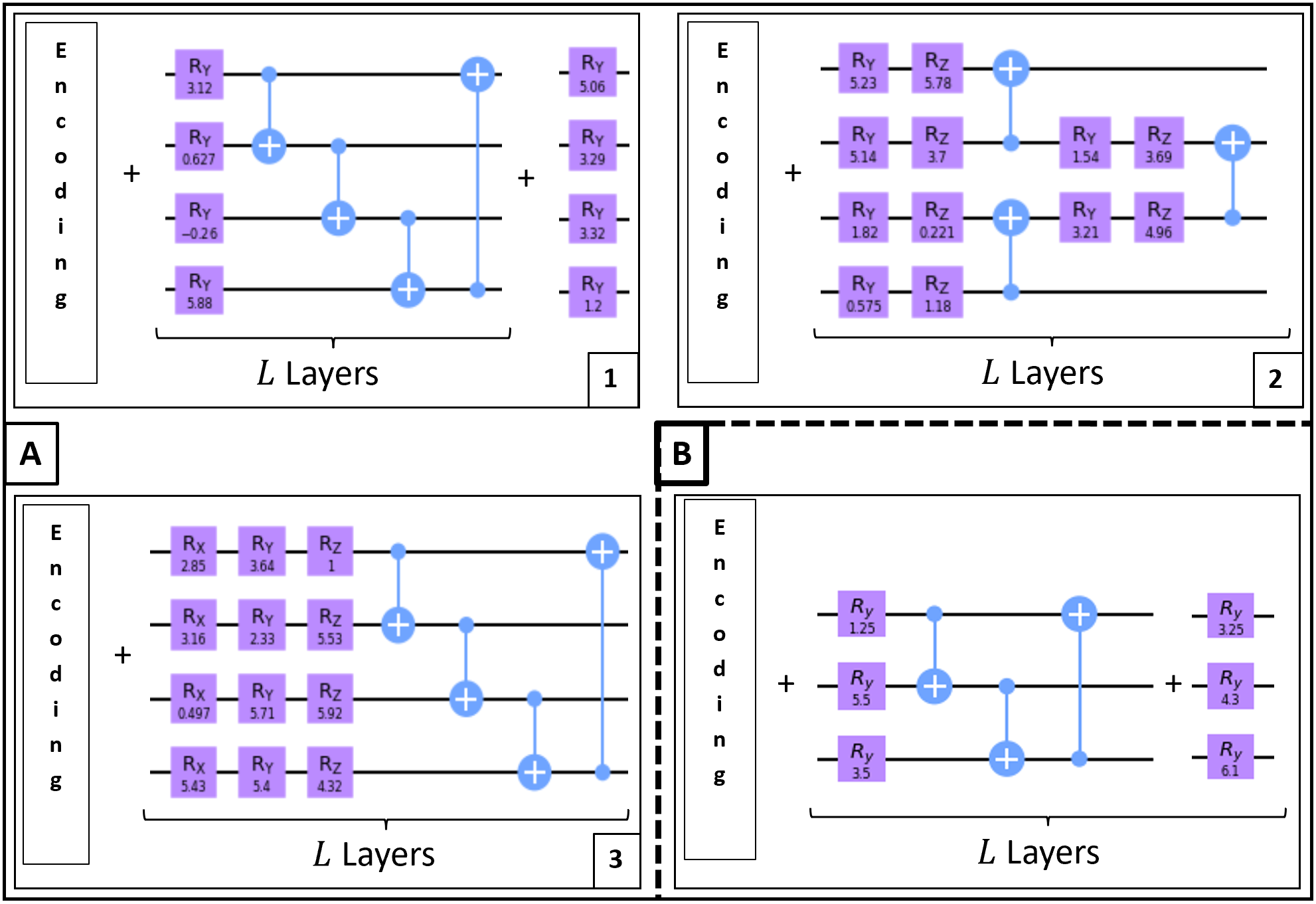}
\caption{\textbf{A}: The 3 Parameterized Quantum Circuits (PQCs) we use for our simulated data compression schemes. The circuits are named Circuit 1, Circuit 2 (from \cite{sim}), and Circuit 3, indicated by the label on the bottom right of each circuit. \textbf{B}: PQC for 3 qubits based on Circuit 1, used for implementing 3 to 2 qubit data compression on an IBMQ device.}
\label{fig:pqc}
\end{figure}

In order to efficiently compress our images into fewer qubits, we quantitatively describe certain properties of the PQC such as performance and resource costs. Using these properties aids us in constructing our circuit with optimal gate sets, which in turn minimize the cost function of the training circuit while limiting computational complexity. First, we look at two circuit descriptors, expressibility and entangling capability, as defined in Sim et al. (2019) \cite{sim}. The authors show expressibility and entangling capability measures for 19 PQCs in their paper \cite{sim}, one of which we use for our simulations (Figure~\ref{fig:pqc}A Circuit 2).

After encoding the data as quantum states (a step we will cover in the following section), we apply the parameterized circuit that best ensures that the data encoding preserves all features of the original data. Thus, an ideal PQC must have the ability to generate states that are well representative of the full Hilbert space. For a single qubit, this means that the PQC must have the ability to explore the full Bloch Sphere. The quantity that measures this property is called the expressibility 
$\epsilon$ of a circuit \cite{sim}. 
Expressibility is measured by generating a sampled distribution of state fidelities of the parameterized circuit, and a sampled distribution using Haar random states (uniform distribution of random states) \cite{kus}. We then measure the Kullback-Leibler (KL) divergence \cite{kldiv} between the two distributions, given by equation (\ref{epsilon}). The KL divergence, or relative entropy, measures how two probability distributions, $P$ and $Q$, differ from one another, given by equation \ref{DKL}. A divergence measure of $D_{KL}=0$ implies the two distributions are the same. Hence, the smaller the KL divergence, or $\epsilon$, the more expressible the parameterized circuit. 

In equation (\ref{epsilon}), $P_{PQC}(F;\theta)$ is the probability distribution from fidelities resulting from sampling states from the parameterized circuit; $F$ is the fidelity, calculated using sampled parameters $\vec{\theta}$. $P_{Haar}(F)$ is the probability distribution from fidelities resulting from sampling states from Haar random states.
\begin{eqnarray}                 
	\epsilon = D_{KL}(P_{PQC}(F;\theta) || P_{Haar}(F)). \label{epsilon}
	\\D_{KL}(P||Q) = \sum_{x}{P(x) \, log_2\frac{P(x)}{Q(x)}}. \label{DKL}
\end{eqnarray}

For the entangling capability measurement, the Meyer-Wallach (MW) measure \cite{meyer} is used, a single scalar measurement for pure-state entanglement \cite{love}. The entanglement of each individual qubit is measured with the remaining qubits in the system. One limitation with this measure is that it cannot distinguish between entangled states that are fully inseparable, and entangled states that can be separated into subsystems \cite{love}. However, in this study, we are measuring the global entanglement of the PQCs, and the ease of computation and scalability \cite{sim} of the MW measure makes it fitting for our experiments. In the Brennen form \cite{brennen}, this distance measure is written as

\begin{eqnarray}                 
	Q(\ket{\psi}) = \frac{1}{n} \sum_{k=1}^{n}2(1-Tr[\rho_k^{2}]) \label{MW},
\end{eqnarray} 

where $Tr[\rho_k^{2}]$ traces out all but the one-qubit reduced density matrix of the $k$th qubit. For set S of sampled circuit parameters $\theta_{i}$, the entangling capability $E$ \cite{sim} is then given by:

\begin{eqnarray}                 
	E = \frac{1}{|S|} \sum_{\theta_{i}\in S}Q(\ket{\psi_{\theta_{i}}}) \label{Ent}.
\end{eqnarray} 

For this study, we examine three different PQCs to implement into the quantum subroutine. Note that in principle it is possible to build an arbitrarily complex PQC capable of achieving lossless or near-lossless compression. However, complex PQCs with large gate depths are detrimental to quantum data compression in the Noisy Intermediate Scale Quantum (NISQ) era \cite{preskill}. Hence, we aim to build PQCs with certain characteristics in mind that make quantum data compression more achievable.

Along with expressibility and entangling capability, we describe the performance of our circuits by examining the resource costs for each run. Figure~\ref{fig:pqc}A shows the three PQCs used for simulations in this study, each composed of 4 qubits.
Table 1 shows the resource costs per circuit for the general case of $n$ qubits and $L$ layers for each of the three circuits. Table 2 shows the expressibility and entangling capability of each PQC studied in this paper with layers $L = 3$. 
Intuitively, increasing the number of layers $L$ of a PQC should improve the performance of the compression, consequently also requiring more computational time while training the circuit. However, for expressibility and entangling capability, the circuits can reach a saturation point, whereby increasing the number of layers of the PQC does not improve the metrics or the PQC's performance. In this study, we look at two PQCs that are more resource intensive and better suited for simulations (PQCs 2 and 3 in Fig~\ref{fig:pqc}A), and one PQC that can be applicable on NISQ devices for small input sizes (PQC 1 in Fig~\ref{fig:pqc}A). Fig~\ref{fig:pqc}B shows this implementation for 3 qubits, run on an IBMQ device \cite{ibmq}.

\subsection{Classical Optimizer}
\begin{figure*}[th]
\includegraphics[width=0.85\textwidth]{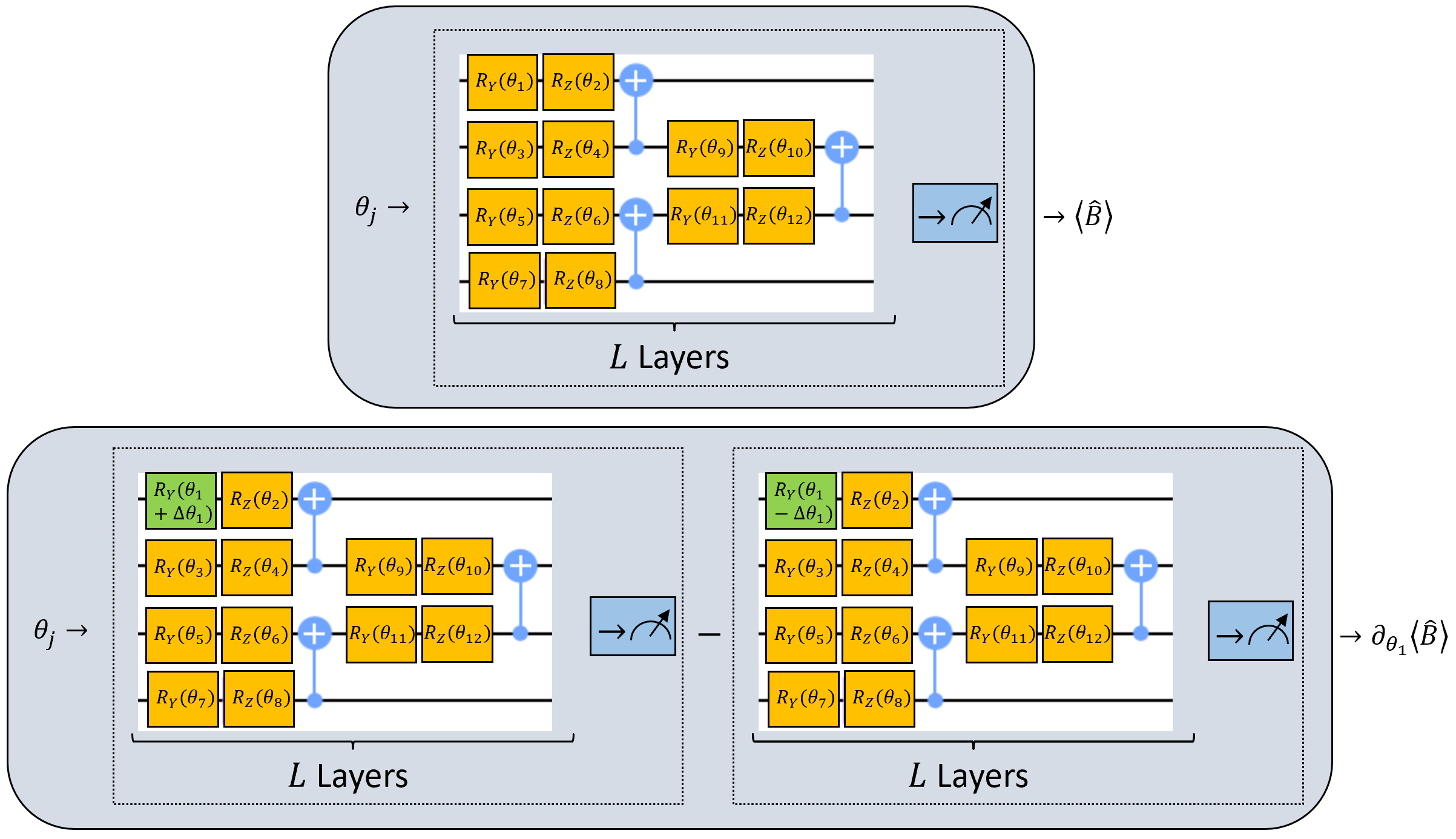}
\caption{Schematic for the Analytical Quantum Gradient Descent: The top figure is one run of the PQC with output observable $\langle\hat{B}\rangle$, and the bottom two figures shows calculation of the gradient of the observable by offsetting each parameter by $\pm\theta\cdot\Delta\theta_j$.}
\label{fig:aqgd}
\end{figure*}

It is important to understand how a classical optimizer works in order to compute the run time of our algorithm. We calculate the number of times our circuit runs per epoch by examining the Analytical Quantum Gradient Descent (AQGD), an IBM Qiskit component that performs gradient descent optimization \cite{mitarai, schuld, qiskitdoc}. One epoch involves running every image in the training dataset once through the compression circuit. After an image is run, the optimizer takes the output observable, $\langle\hat{B}\rangle$, and separately conducts two more runs of the circuit for each parameter in order to calculate the gradient. For each parameter $\theta_j$, the circuit is run once with a positive shift $\theta_j+\Delta\theta_{j}$, giving the output $\langle\hat{B}\rangle^{+}_{j}$, and another run with a negative shift $\theta_j-\Delta\theta_{j}$, giving the output $\langle\hat{B}\rangle^{-}_{j}$, as shown in Figure~\ref{fig:aqgd} ($\Delta\theta=\theta_{j}\cdot\pi/2$). The gradient of the observable $\langle\hat{B}\rangle$ is then given by equation (\ref{gradient}):

\begin{eqnarray}                 
	\frac{\partial\langle\hat{B}\rangle}{\partial\theta_{j}} = \frac{\langle\hat{B}\rangle^{+}_{j} - \langle\hat{B}\rangle^{-}_{j}}{2}. \label{gradient}
\end{eqnarray} 

\begin{eqnarray}                 
	 \langle\hat{B}\rangle' = \langle\hat{B}\rangle - \eta \cdot \nabla \langle\hat{B}\rangle .  \label{gradient_descent_max}
\end{eqnarray}

Calculating equation (\ref{gradient}) for all parameters gives us the total gradient $\nabla\langle\hat{B}\rangle$. We calculate the next step in the gradient descent with equation (\ref{gradient_descent_max}), where $\langle\hat{B}\rangle'$ is the new observable, $\langle\hat{B}\rangle$ is the current observable, and $\eta$ is the learning rate. The learning rate is the coefficient of the gradient update, whereby increasing its value results in larger step sizes for the gradient optimization. Once the new observable $\langle\hat{B}\rangle'$ is found, gradient descent is used to find new adjusted parameters $\theta_j'$ used for the next iteration of training, i.e. $\theta_{t+1}-\theta_t \propto -\eta\nabla\langle \hat{B}\rangle'$ \cite{stokes}. In this study, $\langle\hat{B}\rangle$ is the expectation value of the measured ancilla qubit (from the SWAP test), as shown in Figure~\ref{fig:qae-schematic}B.

The total number of times a circuit is run per image involves 1 run of the original circuit, and 2 runs of the circuit per parameter to calculate $\partial\langle\hat{B}\rangle/\partial\theta_{j}$. This operation can be performed for many iterations of the optimizer, adjustable by an AQGD hyper-parameter. So for the full training dataset, the number of times we run the circuit per epoch is given by:
\begin{eqnarray}                 
	N_{\textrm{jobs}} = ((2 \cdot N_{\textrm{params}} + 1) \cdot N_{\textrm{images}}) \cdot N_{\textrm{iter}} \label{njobs},
\end{eqnarray} 
where $N_{\textrm{iter}}$ is the maximum number of iterations of the optimizer, $N_{\textrm{images}}$ is the number of images in the training dataset that the training circuit goes through in one epoch,  and $N_{\textrm{params}}$ is the total number of parameters in the PQC. Here, $N_{\textrm{iter}}$ is equivalent to number of shots (circuit runs + measurement) the optimizer takes to calculate the average cost function in equation (\ref{cost}). For all simulations in this paper, the maximum number of iterations, $N_{\textrm{iter}}$, was set to 10.

\begin {table}
\caption {Circuit Costs}
\begin{center}
\begin{tabular}{|c|c|c|c|}
\hline
\textbf{Circuit} & \vtop{\hbox{\strut \textbf{Number of}}\hbox{\strut \textbf{Parameters}}} & \vtop{\hbox{\strut \textbf{Number of two}}\hbox{\strut \textbf{qubit gates}}} & \textbf{Circuit Depth}\\
\hline
\textbf{1} & $n(L+1)$ & $nL$ & $(n+1)L+1$\\
 \hline
 \textbf{2} & $4(n-1)L$ & $(n-1)L$ & $6L$\\ 
 \hline
 \textbf{3} & $3nL$ & $nL$ & $(n+3)L$\\ 
 \hline
\end{tabular}
\end{center}
\end{table}

\begin {table}
\caption {PQC Descriptors}
\begin{center}
\begin{tabular}{|c|c|c|}
\hline
\textbf{Circuit} & \textbf{Expressibility} ($\mathbf{D_{KL}}$)  & \vtop{\hbox{\strut \textbf{Entangling}}\hbox{\strut \textbf{Capability}}}\\
\hline
\textbf{1 ($n=4, L=3$)} & 0.130 & 0.800 \\
 \hline
 \textbf{2 ($n=4, L=3$)} & 0.008 & 0.743 \\ 
 \hline
 \textbf{3 ($n=4, L=3$)} & 0.005 & 0.826 \\ 
 \hline
\end{tabular}
\end{center}
\end{table}

\section{Data Encoding}
\subsection{Data for Simulations}
\begin{figure}
\includegraphics[width=0.5\textwidth]{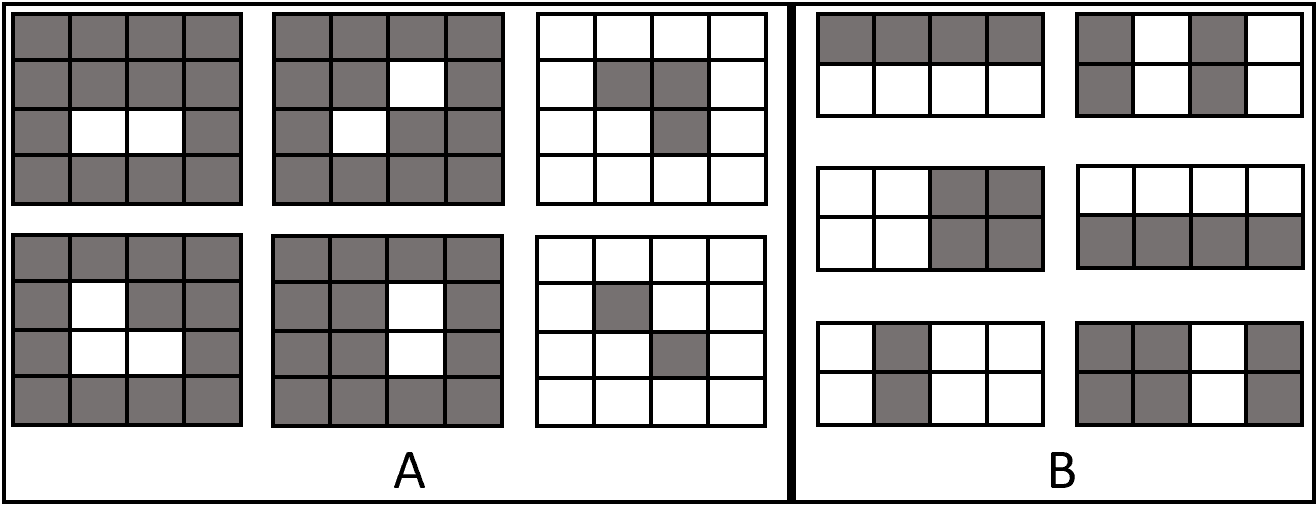}
\caption{
\textbf{A (for simulations)}: 6 sample images from our 4x4 pixelated image dataset. The rows and columns are labeled by $\{00,10,01,11\}$, and $\ket{(c_1,c_2),(r_1,r_2)}$ indicates the pixel label by the two column (row) bits $c_1\,c_2,(r_1\,r_2)$. \textbf{B (for device)}: 6 sample images from the 2x4 bar-and-stripe pattern dataset. Here, $\ket{(c_1,c_2),r_1}$  indicates the pixel label by the two column bits $c_1\,c_2$, and single row bit $r_1$. 
}
\label{fig:picturesque}
\end{figure}

In order to train the network to efficiently compress quantum data, we use a data set consisting of black and white pixelated images following specific rules. For the simulations, we use 4x4 pixel images where all squares touching the border are either black or white, creating a ``frame'' around the 4 central squares, thereby functioning as a single pixel. The number of arrangements in the full dataset is then dependent on the central 4 squares and the frame, yielding a total $2^5$ variations. Figure~\ref{fig:picturesque}A shows 6 examples out of 32 total images. Using 4x4 images (16 pixels) allows us to represent each image as an equal superposition state of 4 qubits, giving us 16 total states for the 16 pixels. The phases on each state encode the different images by the following sign convention: phases on the superposition states go as $(-1)^n$, where $n$ = 0 for a black pixel (positive phase), and $n$ = 1 for a white pixel (negative phase). Phases on the pixels are assigned left to right, and top to bottom, following the convention in Tacchino et al. (2019) \cite{tacchino}. This encoding of pixels creates a dataset of unique linearly independent vectors encoded using 4 qubits, which we then compress down to 3, 2 and 1 qubit in our simulations. The border rule on the dataset ensures a specific pattern for the images, and limits the dataset to size $2^{5}$, instead of $2^{16}$. The density matrices generated from the quantum states are discrete, and allow for efficient and lossless compression to a more continuous latent space. In Figure~\ref{fig:picturesque}A, the top-right image would have the following associated superposition state: $-0.25\ket{0000}-0.25\ket{1000}-0.25\ket{0100}-0.25\ket{1100}-0.25\ket{0010}+0.25\ket{1010}+0.25\ket{0110}-0.25\ket{1110}-0.25\ket{0001}-0.25\ket{1001}+0.25\ket{0101}-0.25\ket{1101}-0.25\ket{0011}-0.25\ket{1011}-0.25\ket{0111}-0.25\ket{1111}$.

Having established our dataset and our methodology for distinguishing different quantum states for each image, we now utilize IBM's Qiskit programming language to produce these states for the simulator. Qiskit's simulator includes a built-in function for complex amplitude initialization, called Initialize \cite{qiskitdoc}. This Initialize function first takes the desired initialized quantum state to the zero state $\ket{0}^{{\otimes}n}$ in the computational basis. The gate sequence that accomplishes this is then implemented backwards after resetting the qubits, taking the $\ket{0}^{{\otimes}n}$ quantum state to the desired initialized state. Hence, Initialize is not a unitary gate, but a resource intensive Qiskit instruction that iteratively disentangles qubits from the register one by one. Since manually assigning control-not and control-phase gates to get the desired state for each 4-qubit image can become cumbersome, we opted for using the Initialize function for our 4-qubit simulations. Future applications with different datasets would require data loading operations unique to those datasets. 

With a total 32 images, we split our dataset with 14 images for training, and 18 images for testing. We augmented the training images by duplicating the dataset, shuffling the list, and picking a small batch of images at a time to train, thereby enabling the training network to train on each image multiple times through one epoch. In total, we trained the network using 42 images with batch size of 7. All simulations were run for 40 epochs with a fixed learning rate of 0.05.

\subsection{Quantum Device Implementation}
Due to limitations in quantum hardware (number of qubits and qubit connectivity), we created a smaller 2x4 pixel image dataset for compression when working with IBMQ. This involved encoding 8 pixels into a 3-qubit superposition state, using the same phase encoding scheme as done for the 4-qubit images. The dataset consists of 10 training images with bar and stripe patterns, where the black/white pixels form horizontal bars, or vertical stripes, but never both. Sample of 6 images are shown in Figure~\ref{fig:picturesque}B. The top-right image would have the following associated superposition state: $0.35355\ket{000}-0.35355\ket{100}+0.35355\ket{010}-0.35355\ket{110}+0.35355\ket{001}-0.35355\ket{101}+0.35355\ket{011}-0.35355\ket{111}$. 
Similar to the 4x4 image dataset, we augmented this dataset to 20 total training images. We used the PQC shown in Figure~\ref{fig:pqc}B, and compressed the 3-qubit superposition state down to 2 qubits. For comparison, this image dataset was compressed with a simulator as well, using the same optimizer and training network hyper-parameters as for the quantum device. For this study, we used the IBMQ `Casablanca' chip, which is one of the IBM Quantum Falcon Processors \cite{ibmq}. To maximize the efficiency of the training network, the data encoding initialization circuit (a combination of CNOT and CPHASE gates), and the parameterized quantum circuit, were transpiled first before training the network using Qiskit's Transpiler \cite{qiskitdoc}.

\section{Results}
Results from our simulations and device experiments are split into two subsections. The first covers the performance and run time of the compression algorithm for the simulated runs. The second subsection goes over the performance and run times of the algorithm for the IBMQ experiment.
\subsection{Simulations}
\subsubsection{Compression Efficiency}
\begin{figure*}
  \centering
  \caption{Simulations: Density Matrices for single test image through different compression ratios and PQCs.}
  \includegraphics[width=\textwidth,height=\textheight,keepaspectratio]{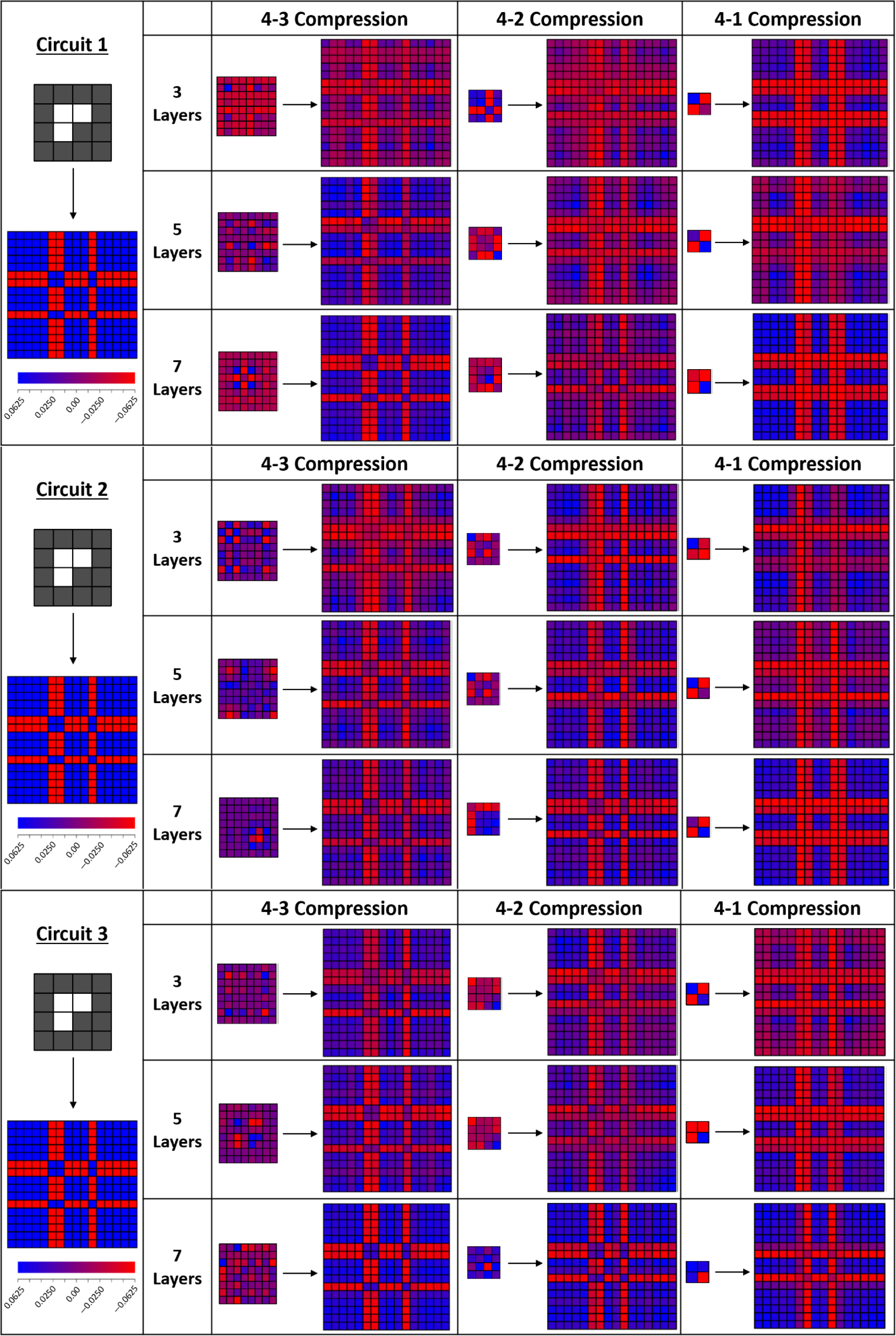}
  \label{fig:full-dm}
\end{figure*}

Visualizing an image undergoing compression and decompression allows us to gauge the performance of our compression algorithm. The original input image and the latent space are visualized via density matrices to show the discrete or continuous nature of the Hilbert space. For the 3 PQCs outlined in Figure~\ref{fig:pqc}A, each one was trained with the Qiskit simulator using layer values of $L=3,5,7$, along with three compression ratios (4 qubits compressed to 3, 2, 1 qubit(s), where the ratio is defined as $n_{\textrm{initial}}/n_{\textrm{compressed}}$), giving a total of 27 full training simulations. Figure~\ref{fig:full-dm} illustrates the full set of simulation results for a single test image, shown as density matrices. The leftmost column shows the example 4x4 image and the associated density matrix visualized as a 16x16 matrix $(\rho = \sum_{i}{p_{i}\ket{\psi}\bra{\psi})}$. The density matrices take on values in the range $\{-0.0625,0.0625\}$, with $\rho_{\textrm{initial}}$ taking discrete values $\pm 1/16$. 
 
The top third section of Figure~\ref{fig:full-dm} shows the simulation results for Circuit 1. The nine cells display the latent space and the decompressed density matrix of the example image for the 3 different compression ratios and 3 PQC depths. The size of the latent space corresponds with the compression ratio; for the 4 to 3 qubit compression, the latent space is an 8x8 matrix. In general, the latent space is $2^{m}$x$2^{m}$, where $m$ is the number of compressed qubits. The forms of the latent space differ for each PQC, regardless of achieving a high fidelity compression. 
In each case, we decompress back to a 16x16 density matrix, with the aim of recreating the original density matrix shown in the leftmost column. The latent space is continuous, while the 16x16 density matrices is discrete. The discrete nature in the initial density matrix can allow us to compress the data down to a continuous mixed state, and potentially without losing information. Theoretically, the data meets the criteria for lossless compression even for the 1 qubit compression since we would have one linearly independent vector compressed down to a latent space size 2. 

So theoretically, there should be no information loss across each compression ratio. In practice, Figure~\ref{fig:full-dm} shows that the greater the compression ratio, the worse the compression algorithm performs. Conversely, increasing the number of layers of the PQC decreases loss of information, and improves compression efficiency. For Circuit 1, 7 layers yields a near lossless compression for the 4 to 3 qubit compression ratio for this particular image. For Circuit 2 with 7 layers, both 4 to 3 and 4 to 2 qubit compressions perform well, while Circuit 3 with 7 layers performs best across the three compression ratios for this image. Ultimately, for any image in the dataset, information loss is proportional to increase in compression, and inversely proportional to number of PQC layers.\par 

\begin{figure}[ht]
\centering
\includegraphics[width=0.5\textwidth]{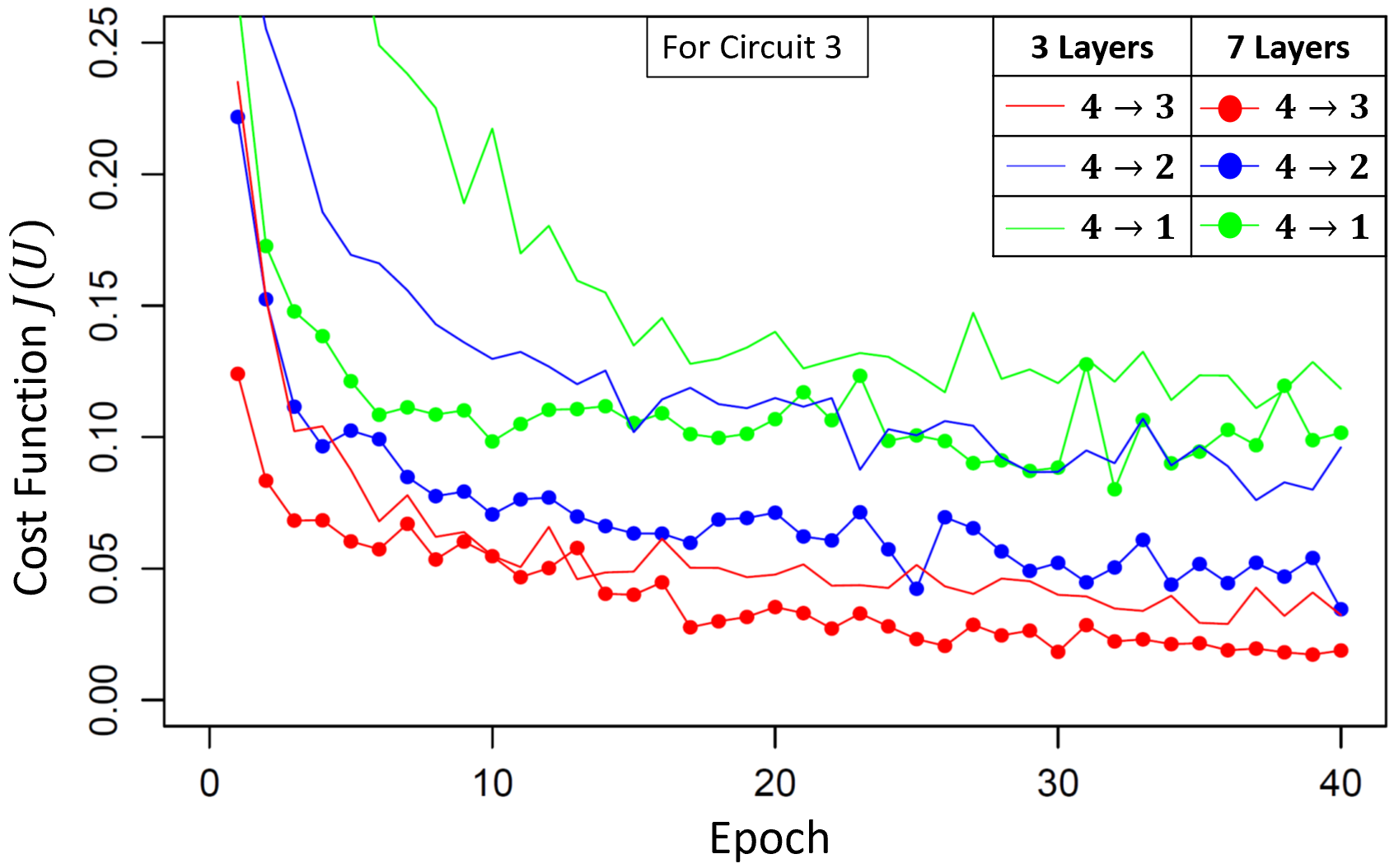}
\caption{Simulations: Loss curves from simulations for Circuit 3 with the three compression ratios, and layers $L=3,7$. We minimize the cost function to approximately 0.02 for 4-to-3 qubit compression with 7 layers.}
\label{fig:losscurveC3}
\end{figure}

This relationship can be seen more clearly in Figure~\ref{fig:losscurveC3} for Circuit 3. The cost function relates to the amount of information lost from data compression via equation (\ref{cost}). Hence, minimizing the cost function is directly related to minimizing information loss, and true lossless compression would see the cost function go down to zero. Figure~\ref{fig:losscurveC3} shows loss curves from simulation of Circuit 3 for the full dataset. At the end of 40 epochs the smallest compression ratio (4 to 3 qubits), with 7 PQC layers, yields the least information loss, minimizing the cost function to 0.02.

\begin{figure*}
\includegraphics[width=\textwidth,height=\textheight,keepaspectratio]{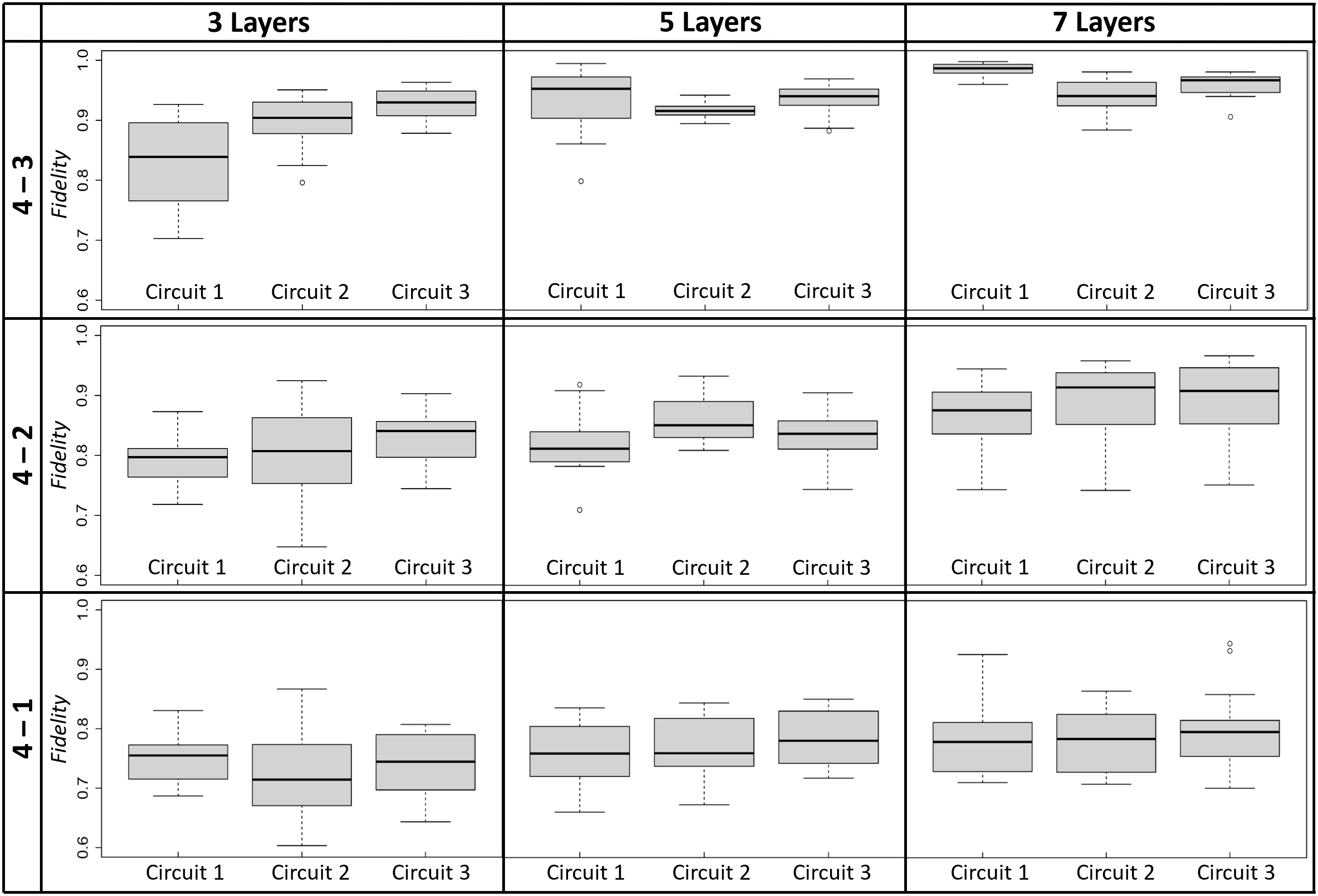}
\caption{Simulations: Fidelity measurements between the original and decompressed density matrices of the full test dataset for the 27 simulations. A score of 1.00 indicates that the original and decompressed image are identical. We show the spread in fidelity scores across the test dataset for each compression ratio, PQC, and number of layers. The circled dots in some of the box plots are marked as outliers (outside 1.5 times the interquartile range above the upper quartile and below the lower quartile).}
\label{fig:boxplot}
\end{figure*}

While Figure~\ref{fig:full-dm} visualizes density matrices for a single image, and provides an insight into how the compression algorithm performs across the 27 simulations, Figure~\ref{fig:boxplot} gives further insight by quantitatively describing how well the PQCs performed on the full test dataset. Figure~\ref{fig:boxplot} plots the fidelity between the original and decompressed images from the full test dataset. Here, we compare the direct results of compression efficiency from simulation per each test image. The box plots show the spread in fidelity measurements across the dataset, with the box containing $50\%$ of the data, and the horizontal black bar as the median value. The highest fidelity measure across all simulations is 0.98, while the lowest is 0.65. For the original density matrix of an image $\rho$, and the decompressed density matrix $\sigma$, the fidelity is calculated using equation (\ref{fidelity}) \cite{chuang}. The maximum fidelity of 1 means the final and initial image is identical, and the compression is lossless.
\begin{eqnarray}                 
	F(\rho, \sigma) \equiv Tr(\sqrt{\rho^{1/2}\sigma\rho^{1/2}}). \label{fidelity}
\end{eqnarray} 

In Figure~\ref{fig:boxplot}, the general trend for PQCs with 3 layers (the leftmost column) shows Circuit 3 performing slightly better than the other two. Since the entangling capability and expressibility measures for Circuits 2 and 3 are more favorable compared to Circuit 1 (Table 2), we see this reflected in the fidelity measures, where Circuit 2 and 3 generally outperform Circuit 1. This can clearly be seen in the simplest case of 4 to 3 qubit compression with 3 layers of PQC (top left of Figure~\ref{fig:boxplot}). This result agrees with Hubregtsen et al. (2021), who found a strong correlation between classification accuracy of their circuit with expressibility of the circuit \cite{hubregtsen}. Even in the general case, Circuit 3 performs slightly better than the other two circuits given that its median fidelity measure is the highest or near highest in all cases. Additionally, aside from the simulation with the smallest compression ratio and the most number of PQC layers (top right of Figure~\ref{fig:boxplot}), Circuit 1 has the largest spread in fidelity measures. Increase in spread of fidelity measures generally corresponds with increase in compression ratios, but not with increase in number of layers, even if the median fidelity score improves with number of PQC layers.

\begin{figure}[ht]
\centering
\includegraphics[width=0.5\textwidth]{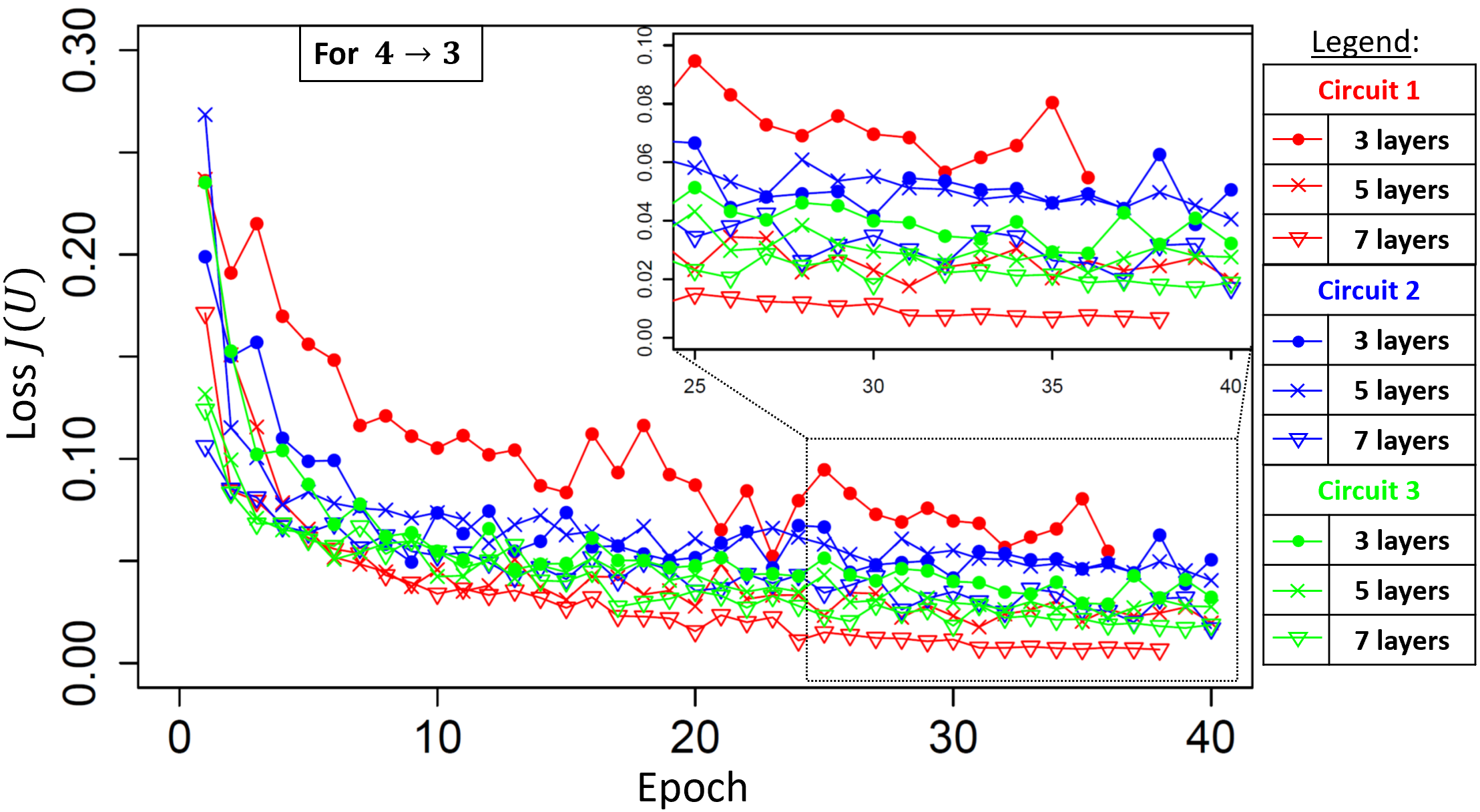}
\caption{Simulations: Loss curves from simulations for all three circuits for the 4-to-3 qubit compression, with layers $L=3,5,7$. The spread in the loss curve is largest with Circuit 1 (red), where increasing the number of layers has the largest effect in minimizing the cost function.}
\label{fig:losscurveall}
\end{figure}

While the increase in number of layers can improve the compression efficiency, each PQC behaves differently to this increase, as evident in Figure~\ref{fig:losscurveall}. The overall compression efficiency of Circuit 1 is most susceptible to change with number of layers, shown as red lines in the plot. Increasing the number of layers for Circuit 2 and 3 (blue and green lines, respectively) does not significantly change the compression efficiency of the algorithm using the simulator on the full test data set, albeit increasing the running time.

\subsubsection{Run Time}
\begin{figure}[ht]
\centering
\includegraphics[width=0.5\textwidth]{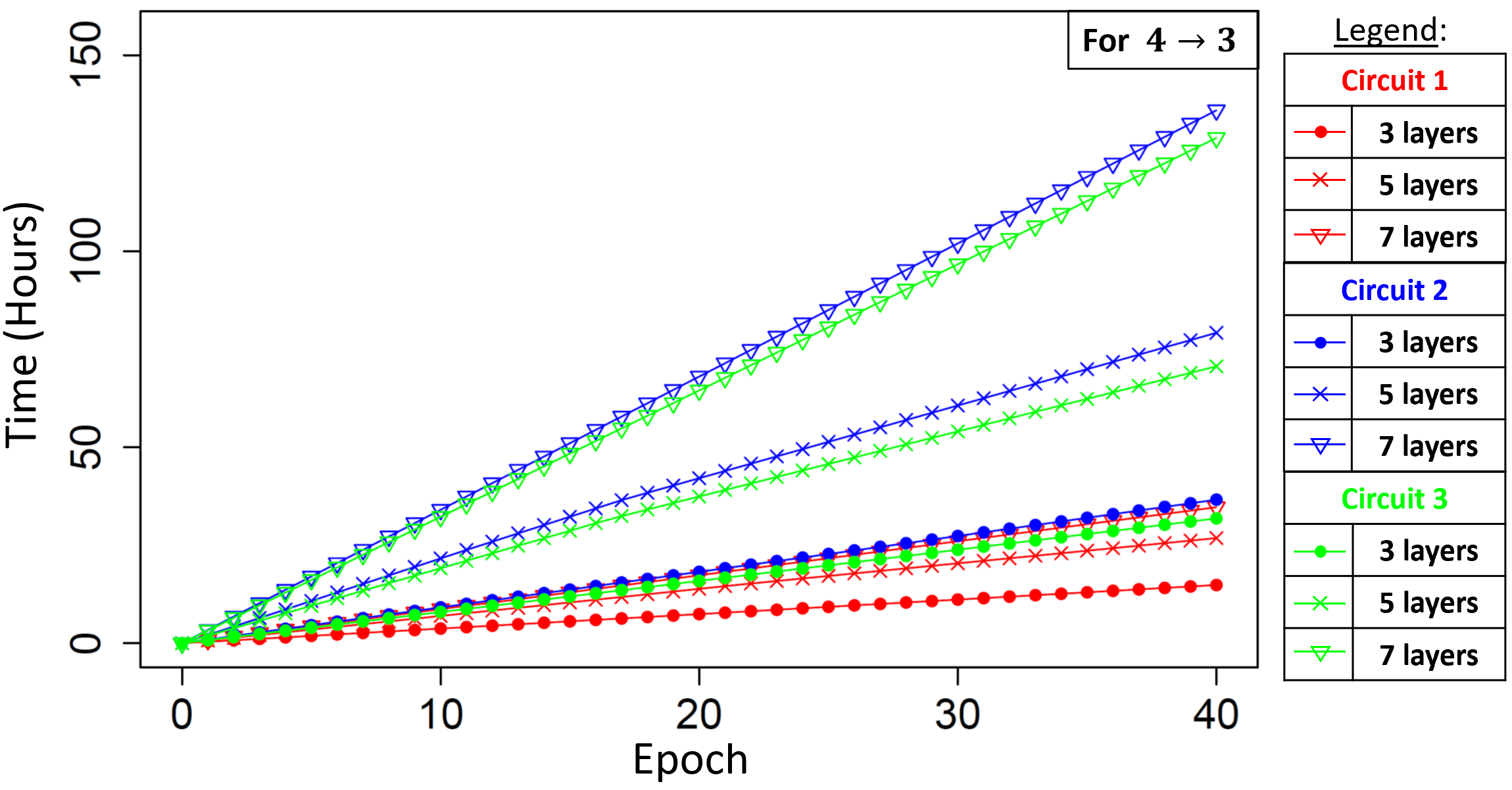}
\caption{Simulations: Total Time (in hours) across 40 epochs for the 4-to-3 qubit compression with the 3 PQCs. Circuit 1 has the smallest spread. Circuit 3 with 3 layers is has a faster training time than Circuit 1 with 7 layers, while Circuit 2 is slowest overall.}
\label{fig:epochtimes}
\end{figure}

\begin {table}
\caption {Simulations: Avg. Total Time per epoch (Hours)}
\begin{center}
\begin{tabular}{|c|c|c|c|}
\hline
\textbf{} & \textbf{Circuit 1} & \textbf{Circuit 2} &  \textbf{Circuit 3}\\
\hline
\textbf{3 Layers} & $0.37$ & $0.91\pm0.01$ & $0.79$\\
 \hline
 \textbf{5 Layers} & $0.67\pm0.03$ & $1.98\pm0.15$ & $1.76\pm0.01$\\ 
 \hline
 \textbf{7 Layers} & $0.87\pm0.01$ & $3.40$ & $3.22$\\ 
 \hline
\end{tabular}
\end{center}
\end{table}

\begin {table}
\caption {IBMQ Quantum Chip: Avg. job time (seconds)}
\begin{center}
\begin{tabular}{|c|c|c|c|}
\hline
\textbf{Create Time} & \textbf{Queue Time} & \textbf{Run Time} & \textbf{Total Time}\\
\hline
0.66 & 46.09 & 20.28 & 68.27 \\
\hline
\end{tabular}
\end{center}
\end{table}

One major focus of this study was to successfully compress quantum data near-losslessly, and quantify how changing the training network hyper-parameters affect information loss and compression efficiency. However, in practice, one must also take into account the running time of the algorithm in question when working with real data. In  Figure~\ref{fig:epochtimes}, for the 4 to 3 qubit compression, we compare simulation data for the total time taken to train the three PQCs across 40 epochs, with 10 iterations of the optimizer. Circuit 1 is the most practical of the three, taking the least time to train and minimize the cost function. It also has the least spread in training time when increasing layers. These characteristics are attributed to its shallow gate depth, which is also reflected in poorer entanglement and expressibility scores. Conversely, Circuit 2 is the slowest circuit across all layers, with the slowest iteration of Circuit 1 being faster than the fastest iteration of Circuit 2.

To study the applicability between the other two circuits, 1 and 3, we examine two competing aspects between the two PQCs: speed vs. performance. The median fidelity measure for Circuit 3 in the the top left box (3 layers) of Figure~\ref{fig:boxplot} is 0.92, while the median fidelity measure for Circuit 1 in the top right box (7 layers) is 0.98. That is approximately a $6\%$ improvement in the median fidelity for Circuit 1. Consequently, the average time per epoch between these two circuits, shown in Table 3, shows Circuit 3 (3 layers) is approximately $9\%$ faster than Circuit 1 (7 layers). So for simulated quantum data compression, to minimize the run time while maximizing performance, Circuit 3 with fewer layers is more favorable. Generally, PQCs with better expressibility and entangling capability measures, like Circuit 3, tend to have higher gate depths, and are only suited to simulated data compression. Provided a $6\%$ increase in information loss can be tolerated by a particular dataset or problem, simulating a more complex PQC with few layers becomes useful. 
Training networks run on quantum devices, however, requires both faster run times and shallower gate depths, with a preference for overall shallower gate depth over number of layers implemented. Hence, Circuit 1's design (shown in Figure~\ref{fig:pqc}B) was chosen to run data compression on the IBMQ quantum device. 

\subsection{IBMQ Device}
\subsubsection{Compression Efficiency}
\begin{figure}[ht]
\centering
\includegraphics[width=0.5\textwidth]{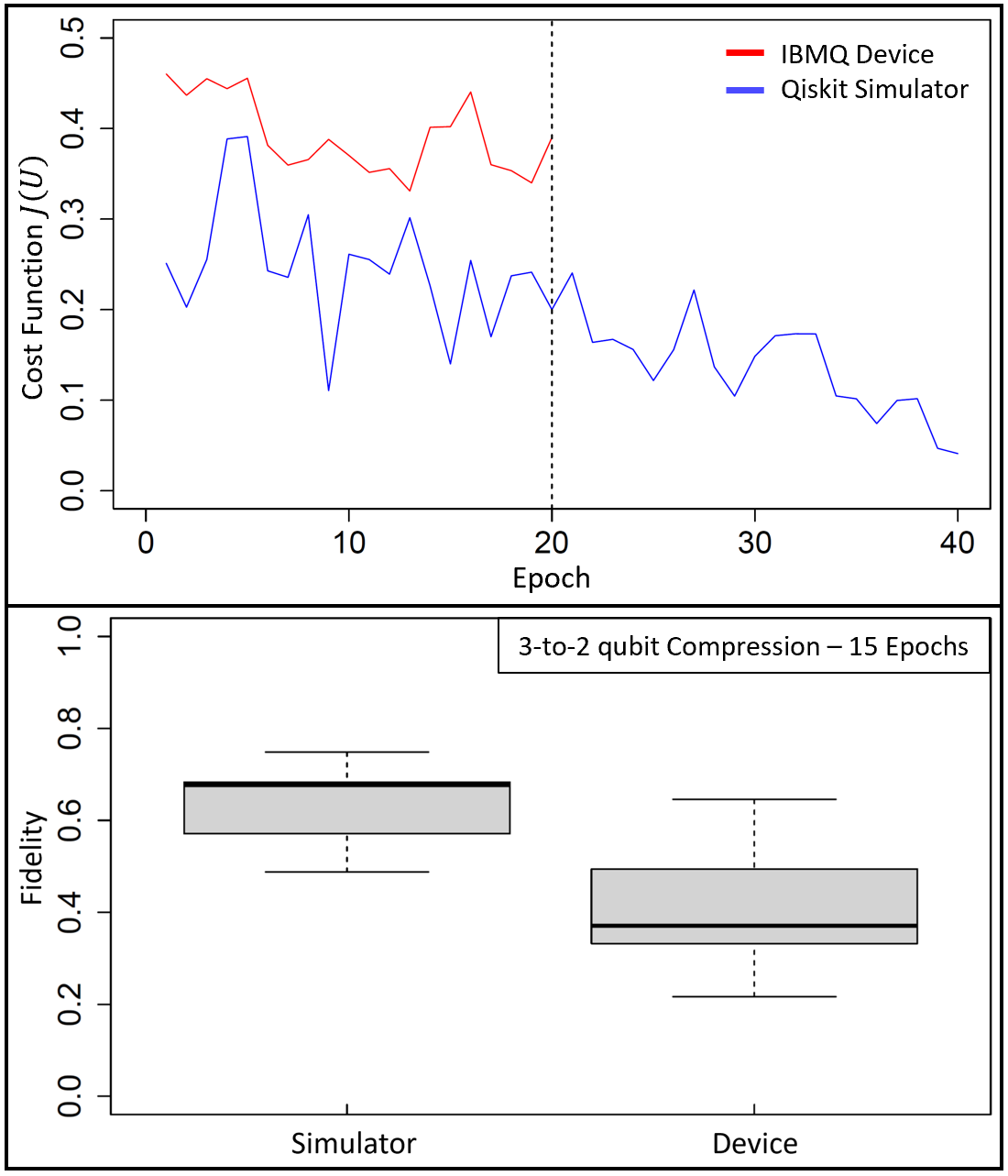}
\caption{\textbf{Top}: Loss curves from IBMQ Device and simulator for 3-to-2 qubit compression with 1 iteration of the optimizer, learning rate of 0.05, and 3 layers of PQC in Figure~\ref{fig:pqc}B. \textbf{Bottom}: Fidelity measures from IBMQ Device and simulator of test dataset with parameters from cost function minima at the end of 20 epochs.}
\label{fig:simdevice}
\end{figure}

Due to the longer training time on the quantum device - primarily due to Queue Time - the network was trained on the device for 20 epochs. Figure~\ref{fig:simdevice} (top) shows the loss curves for the device and simulator, with the vertical line at 20 epochs. We compare the two to show the difference in compression efficiency between the latest superconducting quantum hardware and a simulator. At the end of 20 epochs, the minima of the cost function for the device was over 3 times higher than the cost function for the simulator. The simulator trained the network out to 40 epochs, whereby the cost function was minimized to 0.04, indicating that the 3-to-2 qubit compression of this 2x4 dataset is indeed near-lossless with this particular set of hyper-parameters. Figure~\ref{fig:simdevice} (bottom) shows the fidelity measurements after training with the simulator and the device for 20 epochs. The median fidelity for the IBMQ device is 0.37, while the median fidelity for the simulator is 0.68, around 1.84 times the value.

\subsubsection{Run Time}
For our experiment on IBMQ, we have to calculate the time cost in more detail. This requires a closer look at the number of jobs the training network sends to the device. In our case, three layers of the circuit from Figure~\ref{fig:pqc}B means $N_{\textrm{params}}=12$, and from the training dataset, $N_{\textrm{images}}=20$. So number of jobs sent to device is given by equation (\ref{njobs}) as:
\begin{eqnarray}                 
	N_{j} = ((2 \cdot 12 + 1) \cdot 20) \cdot N_{\textrm{iter}} = 500 \cdot N_{\textrm{iter}}. \label{njobs-num}
\end{eqnarray} 
So for the simplest PQC with 3 layers, as well as our simple image data set compressing 3 qubits down to 2, the number of jobs with just one iteration of the optimizer means 500 jobs were sent to the device per epoch. For 40 epochs, that is 20,000 total jobs sent to the device. Each job is one run of the entire circuit for a set amount of shots (set to a maximum of 8192 shots). The time per job for the training network run on the device, with one iteration of the optimizer, is broken up into several processes, shown in Table 4. Omitting the queue time for the job sent to the device, since it can vary greatly depending on number of users on the device, the average time per job is then 22.18 seconds. For comparison, we ran this dataset through the compression algorithm using a simulator with 1 iteration of the optimizer. The simulator took an average of 7.46 seconds per job (one run of the circuit with 8192 shots). Therefore, the device takes roughly 3 times longer per job compared to the simulator for the same dataset, PQC, and hyper-parameters. 

\section{Summary and Conclusion}
In the Noisy Intermediate Scale Quantum (NISQ) era, qubits are expensive resources, and multi-qubit operations are fragile. By leveraging classical dimensional reduction techniques such as autoencoders, quantum data can be compressed to fewer qubits. Reducing the dimensionality of this problem can allow us to utilize fewer qubits, aiding in more efficient use of resources. However, quantum autoencoders face challenges in efficient data compression even if data sets match the theoretical limit for lossless compression \cite{huang}. Application of quantum data compression needs to address consequences from choice of data, parameterized quantum circuits, and optimizer. In this paper, we successfully demonstrate near-lossless quantum data compression using a simulator. We chose datasets with linearly independent vectors following specific patterns, visualized as black-and-white pixel images. Wavefunctions generated from encoding these images create density matrices with discrete values, allowing for efficient compression down to continuous latent spaces. Three PQCs were tested on the quantum autoencoder framework \cite{romero}, using Qiskit's Analytical Quantum Gradient Descent optimizer. The PQCs were chosen for different gate depths, number of parameters, number of two-qubit gates, and their performance quantified using expressibility and entangling capability measures \cite{sim}. 

Our results show that while all three PQCs manage to efficiently compress the test data set with near-lossless compression using a simulator, their performance was dependent on the compression ratio, number of layers, and running time. For example, Circuit 3 with 3 layers had a $6\%$ higher cost function minimum compared Circuit 1 with 7 layers, but was $9\%$ faster in training time. Only if a $6\%$ increase in information loss can be tolerated, does a more complex PQC with few layers becomes useful, since the speed-up in training can be beneficial. However, for a fixed number of layers, training on a quantum device finds a faster circuit more useful than an efficient one. So although circuits with higher expressibility and entangling capability, like Circuits 2 and 3, generally perform better with larger compression ratios, their complexity leads to noise washing out our results. For devices with limited qubits and connectivity, a simple circuit with smaller gate depth, like Circuit 1, is more NISQ-friendly.

For comparisons of run times and performance of the training circuit between an IBMQ quantum device and the simulator, we used the 3-qubit version of Circuit 1 for the PQC, shown in Figure~\ref{fig:pqc}B. Using a simulator with this PQC on our smaller data set, and one iteration of the optimizer, we compressed from 3 to 2 qubits near-losslessly by minimizing the cost function to 0.04, as seen in Figure~\ref{fig:simdevice} (top). When run on an IBMQ device, the loss curve minima was a little over 3 times larger than the minima for the simulator after 20 epochs. Additionally, the device took three times longer per job compared to the simulator, when omitting queue time. Ideally, increase in total run time due to the job queue can be mitigated by reserving time on the device. We are still at a stage in the NISQ era where not only the noise and limited connectivity affect quantum algorithms, but processes within classical optimizers and data encoding as well.

Ultimately, efficient compression of quantum data on a quantum device will require careful construction of PQCs based on parameters examined in this paper, as well as choice of classical optimizer. In this paper, we examined the performance of our compression algorithm for different PQCs, and IBMQ's Analytical Quantum Gradient Descent optimizer. Future work will involve integrating other gradient descent optimizers into Qiskit that promise a faster convergence, such as the Quantum Natural Gradient proposed by Stokes et al. (2019) \cite{stokes}.
Upon compressing quantum data is efficiently, future work will also involve decompressing latent states into any desired output state by implementing a parameterized circuit for decompression as well. If these compressed states have missing information from the original input, any operation performed using the latent space would compound inaccuracies in the final result. Hence, the goal of this paper to practically achieve lossless compression is important for future applications of quantum data compression.

\begin{acknowledgments}
The views expressed are those of the authors and do not reflect the official guidance or position of the United States Government, the Department of Defense or of the United States Air Force.

The appearance of external hyperlinks does not constitute endorsement by the United States Department of Defense (DoD) of the linked websites, or the information, products, or services contained therein.  The DoD does not exercise any editorial, security, or other control over the information you may find at these locations

\end{acknowledgments}

\nocite{*}

\bibliography{biblio}

\end{document}